# Observation of an exciton crystal in a moiré excitonic insulator

Ruishi Qi[1,2,†], Qize Li[1,3,†], Haleem Kim[1,2], Jiahui Nie[1,2,3], Zuocheng Zhang[1], Ruichen Xia[1], Zhiyuan Cui[1], Jianghan Xiao[1,2,3], Takashi Taniguchi[4], Kenji Watanabe[5], Michael F. Crommie[1,2], Feng Wang[1,2,6,*]

[1] Department of Physics, University of California, Berkeley, CA 94720, USA.

[2] Materials Sciences Division, Lawrence Berkeley National Laboratory, Berkeley, CA 94720, USA.

[3] Graduate Group in Applied Science and Technology, University of California, Berkeley, CA 94720, USA.

[4] Research Center for Materials Nanoarchitectonics, National Institute for Materials Science, 1-1 Namiki, Tsukuba 305-0044, Japan.

[5] Research Center for Functional Materials, National Institute for Materials Science, 1-1 Namiki, Tsukuba 305-0044, Japan.

[6] Kavli Energy NanoScience Institute, University of California Berkeley and Lawrence Berkeley National Laboratory, Berkeley, CA 94720, USA.

† These authors contributed equally.

* To whom correspondence should be addressed: fengwang76@berkeley.edu

**Strong Coulomb interactions can drive electrons to crystallize into a Wigner lattice. Achieving the bosonic analogue – a crystal of excitons – has remained elusive due to their short lifetimes and weaker interactions. Here, we report the observation of a thermodynamically stable exciton crystal in an excitonic insulator coupled to a moiré potential. Using an electron-hole bilayer composed of a monolayer $MoSe_2$ and a $WS_2/WSe_2$ moiré superlattice, we construct a tunable extended Bose-Hubbard model with electrical control over exciton and charge doping in thermal equilibrium. Optical spectroscopy reveals spontaneous crystallization of long-lived excitons at one exciton filling per three moiré sites, evidenced by strong Umklapp scattering peaks in the optical spectrum. Exciton transport measurements further show a pronounced exciton resistance peak at the same filling, consistent with suppressed exciton hopping in a crystalline phase. When doped away from net charge neutrality, this moiré electron-hole bilayer can host new correlated insulating phases where dipolar excitonic insulators form on top of the background of a hole Mott insulator or generalized Wigner crystals in the moiré superlattice. These findings establish moiré excitonic insulators as a versatile platform for realizing correlated crystalline phases of bosons and fermions.**

Quantum matter emerges when particles interact strongly enough to organize themselves into collective phases with no classical counterpart. As a celebrated example, a quantum fluid of electrons can spontaneously break the translational symmetry and crystallize into a Wigner crystal[1] when their Coulomb repulsion dominates over kinetic energy. While such crystalline phases of electrons have been extensively studied in various two-dimensional systems[2–8], the bosonic counterpart – a crystal of excitons – has remained largely unexplored. Realizing it would provide a new platform for exploring strongly correlated bosons, complementing both ultracold atom systems and electronic crystals in solid state systems.

To realize crystalline phases of excitons, it is critical to engineer long lived excitons with strong inter-exciton interactions. Conventionally, excitons are generated in semiconductors through optical excitation, and their interactions can be tuned via quantum confinement or dielectric engineering. Many interesting exciton physics and dynamics have been revealed in examining transient, photoexcited excitons[9–11]. Utilizing moiré superlattices or artificial periodic electrostatic gating, recent studies have enabled trapping and transient ordering of photogenerated excitons, revealing signatures of excitonic Mott states and density waves[12–18]. However, such photoexcited excitons have lifetimes of only pico- to nanoseconds, making them inherently transient and far from thermal equilibrium. It is difficult to realize or obtain definitive signatures of an exciton crystal for such transient nonequilibrium states. Crystallizing excitons in a stable, tunable ground state stands as an elusive goal due to challenges in both achieving long-lived excitons and suppressing their kinetic energy enough for interactions to dominate.

We overcome these barriers by combining a gate-tunable excitonic insulator (EI) in an electron-hole (e-h) bilayer with a moiré potential that suppresses exciton motion. Recent development of e-h bilayers in van der Waals heterostructures[19–24] enabled the realization of long-lived interlayer excitons in the ground state, where layer-separated electrons and holes spontaneously bind into dipolar excitons. With a permanent out-of-plane electric dipole, these interlayer excitons also feature appreciable repulsive dipolar interactions[25]. Introducing a moiré superlattice to one of its layers further quenches the exciton kinetic energy, favoring spontaneous crystallization.

Here we report unambiguous observation of a thermodynamically stable exciton crystal in an e-h bilayer composed of an angle-aligned $WS_2/WSe_2$ moiré superlattice and a monolayer $MoSe_2$ separated by an insulating barrier. Tightly bound dipolar excitons form an EI in the e-h bilayer, whose exciton density can be controlled by electrostatic voltages. We utilize both optical spectroscopy and transport measurements to investigate tunable exciton states in the moiré potential. We show that the dipolar excitons crystallize spontaneously and break the translational symmetry of the underlying moiré lattice when the exciton population is tuned to 1/3 filling (i.e., one exciton per three moiré sites). The exciton crystal gives rise to an Umklapp scattering peak in the optical spectra and a pronounced resistance peak in the exciton transport. In addition, when the bilayer moiré system is doped with both fermionic charges and bosonic excitons, we observe new types of correlated insulating states where dipolar EI is formed on top of charged Mott insulators (MI) or generalized Wigner crystals (GWC).

## Electron-hole bilayers with a moiré potential

Figure 1a illustrates the e-h bilayer structure with a moiré potential. The device consists of an angle-aligned $WS_2/WSe_2$ moiré superlattice as the hole layer and a monolayer $MoSe_2$ as the electron layer. They are individually contacted and separated by an ultrathin (~2 nm) hBN tunneling barrier. The heterostructure is further gated by graphite top gate (TG) and bottom gate (BG) with hBN dielectrics. When no gate voltage is applied, the heterostructure has a type-II band gap of approximately 1.5 eV[21,22], with the conduction band minimum in the $MoSe_2$ layer and the valence band maximum in the $WSe_2$ layer. The $WS_2$ layer only provides a moiré potential and will not be doped. The moiré superlattice, created by the 4% lattice mismatch between $WS_2$ and $WSe_2$, imposes a periodic potential on the hole layer with ~ 8 nm period. The e-h bilayer is in the strong interlayer coupling regime: the interlayer distance of 2 nm is much smaller than the moiré period. The hBN spacer, although only 2 nm thick, provides an insulating barrier with a high tunneling resistance ($10^8$-$10^9$ Ω, Extended Data Fig. 1) ensuring that interlayer tunneling is negligible.

Fig. 1b is an optical image of such a device, D1. We will mainly focus on this device, but consistent data have been reproduced in another device D2 (Extended Data Fig. 2). Fig. 1c shows a typical reflection contrast spectrum of the device in its undoped state, measured at the cryostat base temperature $T = 2$ K. The sharp absorption peak at 1.65 eV comes from the intralayer exciton resonance $X_0$ of the MoSe$_2$ monolayer[26–28]. Similarly, the WS$_2$/WSe$_2$ moiré heterostructure gives rise to multiple moiré exciton peaks of the WSe$_2$ layer within 1.68-1.8 eV, consistent with those observed in previous studies of angle-aligned WS$_2$/WSe$_2$ moiré heterostructures[10,29]. Note that all these peaks originate from intralayer optical transitions within each layer, not from the stable interlayer excitons we will study here, whose oscillator strength is negligible due to e-h separation.

The e-h bilayer structure offers direct electrical control of the band gap and net charge density in the system. The type-II band gap can be closed by a combination of vertical electric field and interlayer bias voltage: The charge gap can be reduced by the vertical electric field ~0.4 V/nm generated by antisymmetric gating $V_{BG}/2 - V_{TG}/2$, and then closed upon application of an interlayer bias voltage $V_B \equiv V_h - V_e$. The symmetric gate voltage $V_G \equiv V_{TG}/2 + V_{BG}/2$ controls the net charge density (i.e., e-h density imbalance)[21–24]. With both $V_G$ and $V_B$, we can control independently the electron density in MoSe$_2$, $n_e$, and the hole density in WS$_2$/WSe$_2$, $n_h$. The exciton resonances presented in Fig. 1c are very sensitive to doped charge carriers – they quickly lose their oscillator strength upon carrier doping in the corresponding TMD layers[22,26–28]. The exciton intensities of different layers therefore provide a direct probe to map out the complete doping configuration of the heterostructure. Fig. 1d and 1e show the reflection contrast at the MoSe$_2$ exciton peak $X_0$ (red dotted line in Fig. 1c) and the lowest energy WS$_2$/WSe$_2$ moiré exciton peak (green dotted line in Fig. 1c), respectively. Based on the intensity maps, we can determine the doping phase boundaries. In Fig. 1d the MoSe$_2$ is intrinsic in the bottom-left region (blue, with strong MoSe$_2$ $X_0$ peak) and becomes electron doped at large $V_G$ or $V_B$ (red, with diminished MoSe$_2$ $X_0$ peak). The white dashed line delineates the electron doping boundary determined by the onset of $X_0$ intensity reduction. In Fig. 1e the WS$_2$/WSe$_2$ layer is intrinsic in the bottom-right region (a large and constant moiré exciton peak) and is hole-doped towards negative $V_G$ or large $V_B$ (reduced moiré exciton peak), with their boundary outlined by the black dashed line. In addition, we notice a region with slightly enhanced moiré exciton intensity in the hole doped WS$_2$/WSe$_2$ layer between the two purple dashed lines. In the same region, the electron doping in Fig. 1d does not depend on $V_B$, i.e. the system is incompressible with respective to the interlayer bias. These are signatures of the MI state in the WS$_2$/WSe$_2$ layer at hole moiré filling factor $n_h/n_0 = 1$ ($n_0 \approx 1.8 \times 10^{12}$cm$^{-2}$ denotes the moiré density with one particle per moiré site)[29].

Figure 1f shows the four-terminal hole resistance ($R_h$) in the WS$_2$/WSe$_2$ moiré layer in a zoomed-in region close to the net charge neutral line. The moiré layer is obviously insulating when there is no hole doping, which includes the band insulator (BI) region at the bottom-middle (with neither electrons nor holes) and the 2D electron gas (2DEG) region at the bottom-right (with electron doping but no holes). Beside these trivial insulating states, there is a triangle-shaped insulating region at finite e-h doping ($V_G \sim 0.22$ V, $V_B \sim 0.75$ V), which is right above the BI phase and follows the net charge neutrality line $n_e = n_h$ (blue dotted lines in Fig. 1f). This is the dipolar EI state similar to that observed in e-h bilayers without a moiré potential[21–24,30,31]. In this phase the single-particle charge gap is still finite, but it is smaller than the interlayer exciton binding energy. Dipolar excitons therefore become energetically favorable and are spontaneously created in the ground state. Such a dipolar EI features spontaneously formed long-lived interlayer excitons[21,22], a large exciton binding energy (~20 meV)[21,22], and perfect Coulomb drag behavior[23,24]. Because electrons and holes all pair into charge-neutral excitons at low temperatures, $R_h$ exhibits insulating behavior. This EI state allows us to investigate the ground-state behavior of excitons in the moiré potential, with exciton density freely adjustable by $V_B$.

## Spectroscopic signatures of exciton crystallization at 1/3 filling

We investigate the EI phase at net charge neutrality. Figure 2a shows the reflection spectrum around the MoSe$_2$ intralayer exciton resonance as a function of $V_B$ following the net charge neutrality line $n_e = n_h$. The energy derivative of reflection contrast is plotted for better visibility of the spectral features. At $V_B < 0.73$ V (left region), the system is a BI with no carrier doping. Increasing $V_B$ will induce a finite interlayer exciton density $n_x$ (= $n_e = n_h$) in the bilayer system. This is signified by the emergence of the intralayer trion peak X$^-$ (also known as the attractive polaron) at lower energy, and the continuous blueshift of the X$_0$ peak[26–28]. Importantly, we find a weaker satellite peak on the high-energy side of the main X$_0$ peak around the exciton filling factor of $n_x/n_0 = 1/3$. The blue curve in Fig. 2b displays the spectrum at 1/3 exciton filling. As a comparison, the magenta line is the spectrum acquired when the system is doped exclusively with electrons at the same density ($n_e = n_0/3, n_h = 0$). The prominent satellite peak is only present when the system is doped with interlayer excitons at 1/3 filling. It is not observable when the system is doped with only electrons at similar densities or at other exciton filling factors. It remains robust up to ~15 K and disappears at higher temperatures (Extended Data Fig. 3).

We attribute the prominent satellite peak to the Umklapp scattering peak in an exciton crystal, as illustrated in Fig. 2c. The optical spectrum normally only probes the MoSe$_2$ X$_0$ energy at zero momentum. With an exciton crystal defining an additional lattice, the MoSe$_2$ X$_0$ dispersion is folded back to the Brillouin zone center, giving rise to an additional absorption peak at a higher energy. Previously satellite peaks have also been reported for electron doped MoSe$_2$ at densities below ~$3 \times 10^{11}$ cm$^{-2}$, and they were identified as the Umklapp scattering from electronic Wigner crystals[2,32]. The Umklapp scattering peak energy is expected to be higher than the main X$_0$ peak by $\Delta E = \frac{h^2 n_x}{\sqrt{3} m}$, which has a value of 9 meV using the interlayer exciton density $n_x = n_0/3 \approx 6 \times 10^{11}$ cm$^{-2}$ and an effective mass of $m = 1.1 m_0$ for the MoSe$_2$ intralayer excitons ($m_0$ denotes the bare electron mass)[2,33]. The experimentally observed energy separation between the main and satellite peaks is about 8 meV, consistent with our interpretation.

The exciton crystal state at 1/3 filling is intriguing. The dipolar excitons feature appreciable dipole-dipole interactions, but their strength is much weaker than the monopole Coulomb interaction between electrons. Therefore, it will be harder to crystallize excitons compared with electron Wigner crystals, and no dipolar exciton crystals have been reported in plain e-h bilayers[21–24,34–36]. Here the moiré superlattice suppresses the exciton kinetic energy and facilitates the formation of exciton crystals at fractional filling. The dipolar exciton state at 1/3 filling can be considered as a "generalized exciton crystal", where dipolar excitons avoid all nearest neighbor moiré site occupations. It breaks the symmetry of the underlying moiré lattice and crystallizes with a period characterized by the $\sqrt{3} \times \sqrt{3}$ lattice vector. We found that the Umklapp scattering peak from the exciton crystal in our moiré e-h bilayer is both stronger and narrower compared with Umklapp scattering peak from electronic Wigner crystal in monolayer MoSe$_2$ at low electron density (Extended Data Fig. 4). It indicates the exciton crystal in the moiré bilayer has a less disordered crystal lattice, presumably owing to the moiré potential in the hole layer constraining possible crystal configurations.

## Exciton transport measurements

We next investigate the exciton transport behavior in such a moiré potential. To perform exciton transport measurements, we fabricate multiple platinum (Pt) contacts for the WS$_2$/WSe$_2$ layer, and few-layer graphite contacts for the MoSe$_2$ layer (Fig. 1b). The transport measurements below are performed in a dilution refrigerator with a base lattice temperature $T = 0.01$ K.

We first use a closed-circuit Coulomb drag experiment to demonstrate the dominance of exciton transport over charge transport in the EI phase. Fig. 3a illustrates the measurement scheme. A small voltage excitation $\Delta V$ is applied between two MoSe$_2$ electrodes to generate a drive current $I_{\text{drive}}$, while two corresponding WS$_2$/WSe$_2$ electrodes are short-circuited by an ammeter to measure the induced drag current $I_{\text{drag}}$. The measured drive and drag currents as functions of $V_B$ along the charge neutral line are plotted in Fig. 3b. At very low $V_B$, neither drive nor drag current can be established in the BI phase (shaded in gray). As $V_B$ is increased into the EI phase (shaded in light blue), an equal amount of drive current and drag current appears. The corresponding drag ratio, defined as $I_{\text{drag}}/I_{\text{drive}}$, reaches unity (Fig. 3c). The perfect drag behavior demonstrates the dominance of exciton transport – all the electrons and holes are bound into interlayer excitons, so the motion of an electron must be accompanied by a hole in the other layer. At higher $V_B$, the drag ratio drops to nearly zero in the electron-hole plasma (EHP) phase (shaded in light yellow). This signifies an interaction-driven exciton Mott transition from tightly bound excitons into two coupled but unbound e-h fluids[21,22,24]. (Although both named after Mott, this exciton Mott transition is not related to and should not be confused with the hole MI states in the WS$_2$/WSe$_2$ moiré layers.) The drive current, $I_{\text{drive}}$, increases by a factor of 10 across the exciton Mott transition due to the emergence of unpaired electrons in the MoSe$_2$ layer. These unpaired electrons reside in the monolayer MoSe$_2$ without a moiré superlattice, which are much more mobile than the dipolar excitons affected by the moiré potential.

Exciton resistance can be estimated from the drag measurement. The exciton resistance $R_x$ is defined as the ratio of the exciton chemical potential drop to the exciton current, analogous to electrical resistance but for a neutral bosonic fluid:

$$R_x = \frac{\Delta V_e - \Delta V_h}{I_x}$$

where $\Delta V_e$ and $\Delta V_h$ are the longitudinal potential drops across the electron layer and the hole layer respectively. In the EI region, the two-terminal exciton resistance can be calculated from the applied excitation voltage and the resulting exciton current. Owing to the perfect drag behavior, the exciton current $I_x$ is equal to $I_{\text{drive}}$. Because the two electrodes of the hole layer are directly shorted through low-resistance contacts, $\Delta V_h$ is negligible; $\Delta V_e$ can be approximated by the applied $\Delta V$ if we neglect the contact voltage drop. We thus get the two-terminal exciton resistance $R_x^{2t} = \Delta V/I_{\text{drive}}$. Fig. 3d provides the measured $R_x^{2t}$ as a function of $V_B$. We observe an overall decreasing trend as the pair density increases, except a peak centered around 1/3 exciton filling. This resistance peak is much higher than the total contact resistance (a few MΩ, Methods), suggesting a strongly insulating phase of the interlayer excitons. We interpret it as the transport signature of an exciton crystal at 1/3 exciton filling, where exciton hopping is suppressed to avoid nearest-neighbor occupation.

To confirm the reliability of this exciton resistance peak, we develop an optical-electrical hybrid four-terminal measurement. The two-terminal resistance inevitably includes complications from contacts. However, it is difficult to perform electrical four-terminal measurements because making reliable low-resistance contacts to n-type semiconducting TMDs at cryogenic temperatures remains very challenging to date. Our optical-electrical hybrid four-terminal method uses electrical bias to drive an exciton current, and optical probes to read out the exciton potential drop inside the heterostructure. It is analogous to a conventional four-terminal measurement, but with optical reflectivity as an effective voltmeter for excitons.

The experimental configuration is illustrated in Fig. 3e. We first apply a common interlayer bias voltage $V_B^0$ to both sides of the device, which establishes a uniform exciton density. This bias condition will be our reference point. On top of $V_B^0$, a small bias difference $\Delta V$ is applied between the two sides to drive an exciton current $I_x$. Specifically, we raise the left side exciton potential by $\alpha \Delta V$ and decrease the right side potential by $(1-\alpha)\Delta V$, where $\alpha$ is a controllable split ratio ($0 \leq \alpha \leq 1$). We limit our focus to the EI phase, so that only exciton current is generated while charge current is irrelevant due to perfect drag. Figure 3f illustrates the spatial varying exciton potential, $V_B(x) - V_B^0$, inside the heterostructure for two different $\alpha$ values. The dark blue line shows the exciton potential distribution as a function of position $x$ at $\alpha = \alpha_1$. With a finite exciton resistance $R_x$ and exciton current $I_x$, the local exciton potential drops linearly with position. The exciton potential reaches zero at $x = x_1$, where $V_B(x_1) = V_B^0$ is not affected by $\Delta V$. Because the reflection intensity from the MoSe$_2$ X$_0$ resonance reflection depends sensitively on the exciton density (as can be seen in Fig. 2a, for example), this special position $x_1$ can be accurately determined using a scanning laser probe, defined by a zero reflection change when $\Delta V$ is turned on and off.

The light blue line in Fig. 3f shows the spatial varying exciton potential at a different value of $\alpha = \alpha_2$. The exciton potential crosses zero at a different position $x = x_2$ that satisfies $V_B(x_2) = V_B^0$ regardless of $\Delta V$. The two different $\alpha$ values correspond to a rigid shift of $V_B$ by $(\alpha_1 - \alpha_2)\Delta V$ at every position. Using the optically determined positions $x_1$ and $x_2$ for $\alpha_1$ and $\alpha_2$, we can determine the exciton potential gradient inside the heterostructure as $(\alpha_1 - \alpha_2)\Delta V/(x_1 - x_2) = \Delta V \frac{d\alpha}{dx}$. The four-terminal exciton resistance $R_x^{4t}$ can then be obtained by dividing the exciton potential drop across the heterostructure by the exciton current,

$$R_x^{4t} = \frac{d\alpha}{dx} \cdot \frac{\Delta V \cdot l}{I_x}$$

where $l$ is the total length of the heterostructure. Because the optical probe is contact-free and samples the device interior, the extracted resistance is unaffected by contact voltage drops. More details and representative raw optical data can be found in Methods and Extended Data Fig. 5.

Fig. 3g shows the optically measured four-terminal exciton resistance $R_x^{4t}$ as a function of $V_B$ (red curve). We observe a clear peak at $n_x/n_0 = 1/3$. The peak is qualitatively consistent with the two-terminal resistance in Fig. 3d, but is sharper and more prominent. It provides direct evidence of the exciton crystal state.

The temperature-dependent exciton resistance at different filling factors is also provided in Fig. 3g and, on a log scale, in its inset. The exciton resistance peak at 1/3 filling remains well defined up to around 10 K, after which the curve becomes featureless. Fitting the exciton resistance to a thermal activation function $\exp(T_0/T)$ gives a temperature scale of $T_0 \approx 17$ K (Extended Data Fig. 6). It is consistent with the temperature where the Umklapp scattering peak disappears, both signifying the melting of the exciton crystal. As discussed later in Fig. 4h and also shown in previous studies[21,22], the interlayer excitons do not ionize into unbound electrons and holes until a much higher temperature. Therefore, at this temperature the exciton crystal thermally melts into an exciton fluid, but individual exciton remains bound together.

**Excitonic insulators coexisting with charge crystals**

Next, we examine the low-temperature phase diagram of the moiré e-h bilayer in the general case where electron and hole densities can be different. Fig. 4a shows the hole-layer resistance $R_\mathrm{h}$ as a function of $V_\mathrm{B}$ and $V_\mathrm{G}$ over a broad voltage range. Apart from trivial insulating regions (BI and 2DEG), we observe many additional correlated insulating phases. The hole MI state at integer filling $n_h/n_0 = 1$ (between two purple dashed lines) is consistent with the optical signature discussed in Fig. 1e. Two narrow insulating lines appear in the hole-doped, electron-undoped region, which are GWC states[3–5] of holes at fractional moiré filling $n_h/n_0 = 1/3$ and $2/3$. The Mott and GWC states are well-known correlated hole states that originate from the strong hole-hole repulsion in the moiré $WS_2/WSe_2$ itself[3–5], without involving electrons in $MoSe_2$ layer.

Interestingly, there are four additional triangle-shaped regions with large $R_\mathrm{h}$ when the bilayer is doped with both electrons and holes. They are more obvious in the drag ratio plot shown in Fig. 4b. While most areas exhibit a negligible drag signal, these four triangular regions have a drag ratio close to unity, indicative of the formation of well-defined interlayer excitons. The most prominent one is the triangular EI region at equal e-h densities (illustrated in Fig. 4g), which shows a perfect drag and has been discussed in previous sections. A second triangular feature near $V_\mathrm{G} \sim -0.05$ V, $V_\mathrm{B} \sim 0.82$ V stems from the hole MI state $n_\mathrm{h} = n_0$ and protrudes along $n_\mathrm{h} = n_\mathrm{e} + n_0$. In this phase, the unity drag ratio and the large $R_h$ value both indicate a finite charge gap that freezes out any appreciable charge current at low temperatures, leaving only excitons mobile. We attribute this state to dipolar excitons coexisting with a hole MI (illustrated in Fig. 4c). In this state, every electron pairs with a hole in the moiré layer, creating a charge gap for the electrons. The remaining excess holes, with a density of exactly $n_0$, form a MI with one hole filled in each moiré site, generating a charge gap for the holes. Neither type of charge can carry net electrical current, but the excitons can hop around to carry an exciton current manifested as the perfect drag behavior.

Two weaker features also appear when the density difference matches fractional moiré fillings 1/3 and 2/3. The drag ratio in these states also approaches unity at low temperatures. They correspond to the coexistence of dipolar excitons and GWCs of additional holes (Fig. 4d-e).

The mixture phases of interlayer excitons and charge crystals are more fragile than the pure EI phase with no excess charges. They only survive for a smaller exciton density range, especially for the EI-GWC mixtures at $n_\mathrm{h} = n_\mathrm{e} + n_0/3$ and $n_\mathrm{h} = n_\mathrm{e} + 2n_0/3$. At low exciton density they can coexist because excitons are charge-neutral and GWCs are sufficiently localized – the screening is weak enough so that they don't melt each other. At higher pair densities many-body interaction effects ionize the interlayer excitons through a quantum phase transition similar to the exciton Mott transition described above. At elevated temperatures, the mixture phases also ionize earlier than the pure EI phase. Fig. 4h shows the temperature-dependent drag ratio for the four excitonic insulating phases in the low exciton density limit. While the drag ratio in the pure EI phase remains nearly perfect (>0.9) up to more than 20 K, the mixture phases only show perfect drag at a few Kelvins. Both facts indicate a smaller exciton binding energy in the mixture phases compared to the pure EI phase due to increased interaction effects from the excess charge.

In summary, we have reported the first realization of a stable exciton crystal in thermal equilibrium. Both spectroscopic and transport evidence unambiguously confirm the crystallization of dipolar excitons at fractional filling of the extended Bose-Hubbard system. We establish excitonic-insulating e-h bilayers coupled with a moiré potential as a versatile platform for realizing and controlling exotic crystalline phases of fermions, bosons, or a tunable mixture of both species, providing an alternate route to ultracold atom systems[37,38].

## Methods

**Device fabrication.** We use a dry-transfer method based on polyethylene terephthalate glycol (PETG) stamps to fabricate the heterostructures. Monolayer $MoSe_2$, monolayer $WS_2$, monolayer $WSe_2$, few-layer graphene and hBN flakes are mechanically exfoliated from bulk crystals onto $SiO_2$/Si substrates. We use 5-8 nm hBN as the gate dielectric, and ~2 nm thin hBN as the interlayer spacer (5-layer hBN for D1, 6-layer hBN for D2). Prior to the stacking process, metal electrodes for the hole layer (7 nm Pt with 3 nm Cr adhesion layer) are defined using photolithography (Durham Magneto Optics, MicroWriter) and electron beam evaporation (Angstrom Engineering) onto a high resistivity $SiO_2$/Si substrate. The graphite gates are patterned using atomic-force-microscope-based anodic oxidation[39] to eliminate undesired current paths in the unmatched monolayer regions. The angle alignment of the $WS_2$/$WSe_2$ moiré bilayer is based on the straight edges of the monolayer flakes. Only flakes with multiple long straight edges forming 60° or 120° angles are selected for assembly. Observation of moiré excitons in the reflection spectrum and GWCs in the transport confirm well-defined moiré patterns.

A 0.5 mm thick clear PETG stamp is employed to pick up the flakes at 65-75 °C. A >100 nm thick hBN is first picked up by the PETG stamp to serve as a protective layer, and then all the following layers are sequentially picked up. The whole stack is then released onto the prepatterned Pt electrodes at ~100 °C, followed by dissolving the PETG in chloroform at room temperature for about one day. Finally, electrodes (5 nm Cr/60-80 nm Au) are defined using photolithography and electron beam evaporation.

Similar to previous works, two carrier reservoir regions with increased interlayer distance are used for better exciton contacts[21–24,30,31]. A large vertical electric field ($\approx$ 0.4 V/nm) is applied to reduce the type-II band gap. The electric field creates a much larger voltage difference in the reservoir regions due to the increased interlayer distance, heavily doping the bilayer system into an EHP that serves as good exciton contacts. At cryogenic temperatures, the contact resistance for the $WS_2$/$WSe_2$ layer is <5 kΩ. The contact resistance for the $MoSe_2$ layer is a few MΩ.

**Reflection spectroscopy measurements.** The reflection spectroscopy and $R_h$ measurements in Figs. 1-2 are performed in an optical cryostat (Quantum Design, OptiCool) with a base temperature of 2 K. The reflection spectroscopy is performed with a supercontinuum laser (Fianium

Femtopower 1060) as the light source. The incident laser power is kept at nanowatt level to minimize heating and photodoping effects. The laser is focused on the sample by a 20× Mitutoyo objective with ~1.5 μm beam diameter. The reflected light is collected after a spectrometer by a CCD camera (Princeton Instruments PIXIS 256e) with 1000 ms exposure time. To minimize the influence of laser fluctuations, another laser beam directly reflected from a silver mirror is simultaneously collected to normalize the sample reflection spectra.

**Coulomb drag measurements.** The Coulomb drag measurement and the hybrid electrical-optical measurement are performed in a dilution refrigerator (Bluefors LD250) with a base lattice temperature of 10 mK, but the electronic temperature could be significantly higher. All the signal wires are filtered at the mixing chamber flange (QDevil) before reaching the sample. The d.c. voltage for the gates is applied with Keithley 2400/2450 source meters or Keithley 2502 picoammeters. In the closed-circuit Coulomb drag measurement, a low frequency a.c. voltage excitation of 5 mV at 17.7 Hz is applied between the two $MoSe_2$ contacts. The drive and drag currents are measured by standard lock-in method (Stanford Research SR865a or SR830).

**Optical-electrical hybrid four-terminal resistance measurement.** Extended Data Fig. 5a shows the detailed circuit diagram. We use the low-frequency lock-in technique to electrically generate the exciton current and optically measure the potential distribution. The a.c. driving voltage $\Delta V$ is distributed to the two sides of the device by a dual-channel arbitrary waveform generator (Siglent SDG6022X). In order to improve the voltage resolution, the output of the arbitrary waveform generator is divided using a 100:1 voltage divider and then supplied to the device. Its two channels are frequency-locked, phase-coupled with a 180-degree phase difference, and programmed to maintain a constant post-divider amplitude sum of $\Delta V = 5$ mV. The resulting driving current is directly measured by standard lock-in method (same as the Coulomb drag measurement).

The optical readout of local potential uses a diode laser whose wavelength is tuned to the $MoSe_2$ main $X_0$ peak by a thermoelectric cooler (gray line in Extended Data Fig. 5b). The laser is focused on the sample with an Olympus PLN 10X objective lens and scanned along the device. Incident laser power is kept below 1 nW to minimize heating and photodoping effects. The light reflected from the device is collected by an avalanche photodiode (Thorlabs APD410A), whose voltage output is analyzed by a lock-in amplifier (Stanford Research SR865a) referenced to the voltage excitation frequency. As shown in Extended Data Fig. 5b, the reflectivity at the laser wavelength depends sensitively on the carrier density. Consequently, the a.c. component of the reflected light intensity is proportional to the local density modulation, and hence to the local potential modulation.

Extended Data Fig. 5c-d displays the measured a.c. reflection contrast signal $\Delta RC$ as a function of beam position $x$ and voltage split ratio $\alpha$. We trace the zero crossing point of the optical signal, yielding a $(x, \alpha)$ relation. Because the analysis uses only the zero crossing of the optical signal, the result does not depend on any proportionality factor between reflectivity and potential. For exciton filling $n_x/n_0 = 1/3$, the position dependence is very strong, indicating a large potential gradient (Extended Data Fig. 5c); for $n_x/n_0 = 0.44$, the position dependence is much weaker due to a smaller exciton resistance (Extended Data Fig. 5d). Although in Fig. 3 we illustrated the measurement scheme with only two $\alpha$ values, the potential gradient inside the device $\Delta V \cdot d\alpha/dx$ is actually obtained by a linear fit of all $(x, \alpha)$ points in the heterostructure. The $(x, \alpha)$ relation at the zero crossing point is provided in Extended Data Fig. 5e at different exciton filling factors.

# References

1. E. Wigner. On the Interaction of Electrons in Metals. *Phys. Rev.* **46**, 1002–1011 (1934).

2. T. Smoleński, P.E. Dolgirev, C. Kuhlenkamp, A. Popert, Y. Shimazaki, P. Back, X. Lu, M. Kroner, K. Watanabe, T. Taniguchi, I. Esterlis, E. Demler & A. Imamoğlu. Signatures of Wigner crystal of electrons in a monolayer semiconductor. *Nature* **595**, 53–57 (2021).

3. H. Li, S. Li, E.C. Regan, D. Wang, W. Zhao, S. Kahn, K. Yumigeta, M. Blei, T. Taniguchi, K. Watanabe, S. Tongay, A. Zettl, M.F. Crommie & F. Wang. Imaging two-dimensional generalized Wigner crystals. *Nature* **597**, 650–654 (2021).

4. E.C. Regan, D. Wang, C. Jin, M.I. Bakti Utama, B. Gao, X. Wei, S. Zhao, W. Zhao, Z. Zhang, K. Yumigeta, M. Blei, J.D. Carlström, K. Watanabe, T. Taniguchi, S. Tongay, M. Crommie, A. Zettl & F. Wang. Mott and generalized Wigner crystal states in WSe2/WS2 moiré superlattices. *Nature* **579**, 359–363 (2020).

5. Y. Xu, S. Liu, D.A. Rhodes, K. Watanabe, T. Taniguchi, J. Hone, V. Elser, K.F. Mak & J. Shan. Correlated insulating states at fractional fillings of moiré superlattices. *Nature* **587**, 214–218 (2020).

6. Y.-C. Tsui, M. He, Y. Hu, E. Lake, T. Wang, K. Watanabe, T. Taniguchi, M.P. Zaletel & A. Yazdani. Direct observation of a magnetic-field-induced Wigner crystal. *Nature* **628**, 287–292 (2024).

7. Z. Xiang, H. Li, J. Xiao, M.H. Naik, Z. Ge, Z. He, S. Chen, J. Nie, S. Li, Y. Jiang, R. Sailus, R. Banerjee, T. Taniguchi, K. Watanabe, S. Tongay, S.G. Louie, M.F. Crommie & F. Wang. Imaging quantum melting in a disordered 2D Wigner solid. *Science* **388**, 736–740 (2025).

8. R.S. Crandall & R. Williams. Crystallization of electrons on the surface of liquid helium. *Phys. Lett. A* **34**, 404–405 (1971).

9. L.V. Butov, C.W. Lai, A.L. Ivanov, A.C. Gossard & D.S. Chemla. Towards Bose–Einstein condensation of excitons in potential traps. *Nature* **417**, 47–52 (2002).

10. C. Jin, E.C. Regan, A. Yan, M. Iqbal Bakti Utama, D. Wang, S. Zhao, Y. Qin, S. Yang, Z. Zheng, S. Shi, K. Watanabe, T. Taniguchi, S. Tongay, A. Zettl & F. Wang. Observation of moiré excitons in WSe2/WS2 heterostructure superlattices. *Nature* **567**, 76–80 (2019).

11. E.C. Regan, D. Wang, E.Y. Paik, Y. Zeng, L. Zhang, J. Zhu, A.H. MacDonald, H. Deng & F. Wang. Emerging exciton physics in transition metal dichalcogenide heterobilayers. *Nat. Rev. Mater.* **7**, 778–795 (2022).

12. C. Lagoin, U. Bhattacharya, T. Grass, R.W. Chhajlany, T. Salamon, K. Baldwin, L. Pfeiffer, M. Lewenstein, M. Holzmann & F. Dubin. Extended Bose–Hubbard model with dipolar excitons. *Nature* **609**, 485–489 (2022).

13. Y. Bai, Y. Li, S. Liu, Y. Guo, J. Pack, J. Wang, C.R. Dean, J. Hone & X. Zhu. Evidence for Exciton Crystals in a 2D Semiconductor Heterotrilayer. *Nano Lett.* **23**, 11621–11629 (2023).

14. Y. Zeng, Z. Xia, R. Dery, K. Watanabe, T. Taniguchi, J. Shan & K.F. Mak. Exciton density waves in Coulomb-coupled dual moiré lattices. *Nat. Mater.* **22**, 175–179 (2023).

15. R. Xiong, J.H. Nie, S.L. Brantly, P. Hays, R. Sailus, K. Watanabe, T. Taniguchi, S. Tongay & C. Jin. Correlated insulator of excitons in WSe2/WS2 moiré superlattices. *Science* **380**, 860–864 (2023).

16. B. Gao, D.G. Suárez-Forero, S. Sarkar, T.-S. Huang, D. Session, M.J. Mehrabad, R. Ni, M. Xie, P. Upadhyay, J. Vannucci, S. Mittal, K. Watanabe, T. Taniguchi, A. Imamoglu, Y. Zhou & M. Hafezi. Excitonic Mott insulator in a Bose-Fermi-Hubbard system of moiré WS2/WSe2 heterobilayer. *Nat. Commun.* **15**, 2305 (2024).

17. Z. Lian, Y. Meng, L. Ma, I. Maity, L. Yan, Q. Wu, X. Huang, D. Chen, X. Chen, X. Chen, M. Blei, T. Taniguchi, K. Watanabe, S. Tongay, J. Lischner, Y.-T. Cui & S.-F. Shi. Valley-polarized excitonic Mott insulator in WS2/WSe2 moiré superlattice. *Nat. Phys.* **20**, 34–39 (2024).

18. C. Lagoin, S. Suffit, K. Baldwin, L. Pfeiffer & F. Dubin. Mott insulator of strongly interacting two-dimensional semiconductor excitons. *Nat. Phys.* **18**, 149–153 (2022).

19. Z. Zhang, E.C. Regan, D. Wang, W. Zhao, S. Wang, M. Sayyad, K. Yumigeta, K. Watanabe, T. Taniguchi, S. Tongay, M. Crommie, A. Zettl, M.P. Zaletel & F. Wang. Correlated interlayer exciton insulator in heterostructures of monolayer WSe2 and moiré WS2/WSe2. *Nat. Phys.* 1–7 (2022) doi:10.1038/s41567-022-01702-z.

20. J. Gu, L. Ma, S. Liu, K. Watanabe, T. Taniguchi, J.C. Hone, J. Shan & K.F. Mak. Dipolar excitonic insulator in a moiré lattice. *Nat. Phys.* **18**, 395–400 (2022).

21. L. Ma, P.X. Nguyen, Z. Wang, Y. Zeng, K. Watanabe, T. Taniguchi, A.H. MacDonald, K.F. Mak & J. Shan. Strongly correlated excitonic insulator in atomic double layers. *Nature* **598**, 585–589 (2021).

22. R. Qi, A.Y. Joe, Z. Zhang, Y. Zeng, T. Zheng, Q. Feng, J. Xie, E. Regan, Z. Lu, T. Taniguchi, K. Watanabe, S. Tongay, M.F. Crommie, A.H. MacDonald & F. Wang. Thermodynamic behavior of correlated electron-hole fluids in van der Waals heterostructures. *Nat. Commun.* **14**, 8264 (2023).

23. R. Qi, A.Y. Joe, Z. Zhang, J. Xie, Q. Feng, Z. Lu, Z. Wang, T. Taniguchi, K. Watanabe, S. Tongay & F. Wang. Perfect Coulomb drag and exciton transport in an excitonic insulator. *Science* **388**, 278–283 (2025).

24. P.X. Nguyen, L. Ma, R. Chaturvedi, K. Watanabe, T. Taniguchi, J. Shan & K.F. Mak. Perfect Coulomb drag in a dipolar excitonic insulator. *Science* **388**, 274–278 (2025).

25. Y.N. Joglekar, A.V. Balatsky & S. Das Sarma. Wigner supersolid of excitons in electron-hole bilayers. *Phys. Rev. B* **74**, 233302 (2006).

26. K.F. Mak, K. He, C. Lee, G.H. Lee, J. Hone, T.F. Heinz & J. Shan. Tightly bound trions in monolayer MoS2. *Nat. Mater.* **12**, 207–211 (2013).

27. G. Scuri, Y. Zhou, A.A. High, D.S. Wild, C. Shu, K. De Greve, L.A. Jauregui, T. Taniguchi, K. Watanabe, P. Kim, M.D. Lukin & H. Park. Large Excitonic Reflectivity of Monolayer MoSe2 Encapsulated in Hexagonal Boron Nitride. *Phys. Rev. Lett.* **120**, 037402 (2018).

28. J.S. Ross, S. Wu, H. Yu, N.J. Ghimire, A.M. Jones, G. Aivazian, J. Yan, D.G. Mandrus, D. Xiao, W. Yao & X. Xu. Electrical control of neutral and charged excitons in a monolayer semiconductor. *Nat. Commun.* **4**, 1474 (2013).

29. Y. Tang, L. Li, T. Li, Y. Xu, S. Liu, K. Barmak, K. Watanabe, T. Taniguchi, A.H. MacDonald, J. Shan & K.F. Mak. Simulation of Hubbard model physics in WSe2/WS2 moiré superlattices. *Nature* **579**, 353–358 (2020).

30. P.X. Nguyen, R. Chaturvedi, B. Zou, K. Watanabe, T. Taniguchi, A.H. MacDonald, K.F. Mak & J. Shan. Quantum oscillations in a dipolar excitonic insulator. Preprint at https://doi.org/10.48550/arXiv.2501.17829 (2025).

31. R. Qi, Q. Li, Z. Zhang, Z. Cui, B. Zou, H. Kim, C. Sanborn, S. Chen, J. Xie, T. Taniguchi, K. Watanabe, M.F. Crommie, A.H. MacDonald & F. Wang. Competition between excitonic insulators and quantum Hall states in correlated electron-hole bilayers. Preprint at https://doi.org/10.48550/arXiv.2501.18168 (2025).

32. J. Sung, J. Wang, I. Esterlis, P.A. Volkov, G. Scuri, Y. Zhou, E. Brutschea, T. Taniguchi, K. Watanabe, Y. Yang, M.A. Morales, S. Zhang, A.J. Millis, M.D. Lukin, P. Kim, E. Demler & H. Park. An electronic microemulsion phase emerging from a quantum crystal-to-liquid transition. *Nat. Phys.* **21**, 437–443 (2025).

33. J. Hong, M. Koshino, R. Senga, T. Pichler, H. Xu & K. Suenaga. Deciphering the Intense Postgap Absorptions of Monolayer Transition Metal Dichalcogenides. *ACS Nano* **15**, 7783–7789 (2021).

34. L. Du, X. Li, W. Lou, G. Sullivan, K. Chang, J. Kono & R.-R. Du. Evidence for a topological excitonic insulator in InAs/GaSb bilayers. *Nat. Commun.* **8**, 1971 (2017).

35. J.I.A. Li, T. Taniguchi, K. Watanabe, J. Hone & C.R. Dean. Excitonic superfluid phase in double bilayer graphene. *Nat. Phys.* **13**, 751–755 (2017).

36. X. Liu, K. Watanabe, T. Taniguchi, B.I. Halperin & P. Kim. Quantum Hall drag of exciton condensate in graphene. *Nat. Phys.* **13**, 746–750 (2017).

37. D. Jaksch, C. Bruder, J.I. Cirac, C.W. Gardiner & P. Zoller. Cold Bosonic Atoms in Optical Lattices. *Phys. Rev. Lett.* **81**, 3108–3111 (1998).

38. M. Greiner, O. Mandel, T. Esslinger, T.W. Hänsch & I. Bloch. Quantum phase transition from a superfluid to a Mott insulator in a gas of ultracold atoms. *Nature* **415**, 39–44 (2002).

39. H. Li, Z. Ying, B. Lyu, A. Deng, L. Wang, T. Taniguchi, K. Watanabe & Z. Shi. Electrode-Free Anodic Oxidation Nanolithography of Low-Dimensional Materials. *Nano Lett.* **18**, 8011–8015 (2018).

## Data availability

The data that support the findings of this study are available from the corresponding author upon request.

## Acknowledgements

We thank Dr. Yubo Yang and Dr. Shiwei Zhang for insightful discussions. The van der Waals heterostructure fabrication and electrical characterization were supported by the U.S. Department of Energy, Office of Science, Office of Basic Energy Sciences, Materials Sciences and Engineering Division under contract no. DE-AC02-05-CH11231 (van der Waals heterostructures program, KCWF16). The optical-electrical four-terminal exciton transport was supported by the AFOSR award FA9550-23-1-0246. K.W. and T.T. acknowledge support from the JSPS KAKENHI (Grant Numbers 21H05233 and 23H02052) and World Premier International Research Center Initiative (WPI), MEXT, Japan.

## Author contributions

F.W. and R.Q. conceived the research. Q.L, R.Q., J.N., R.X., Z.C. and J.X. fabricated the devices. R.Q., Q.L. and H.K. performed the optical measurements. R.Q. and Z.Z. performed the transport measurements. R.Q., Q.L., M.F.C. and F.W. analyzed the data. K.W. and T.T. grew hBN crystals. All authors discussed the results and wrote the manuscript.

## Competing interests

The authors declare no competing interests.

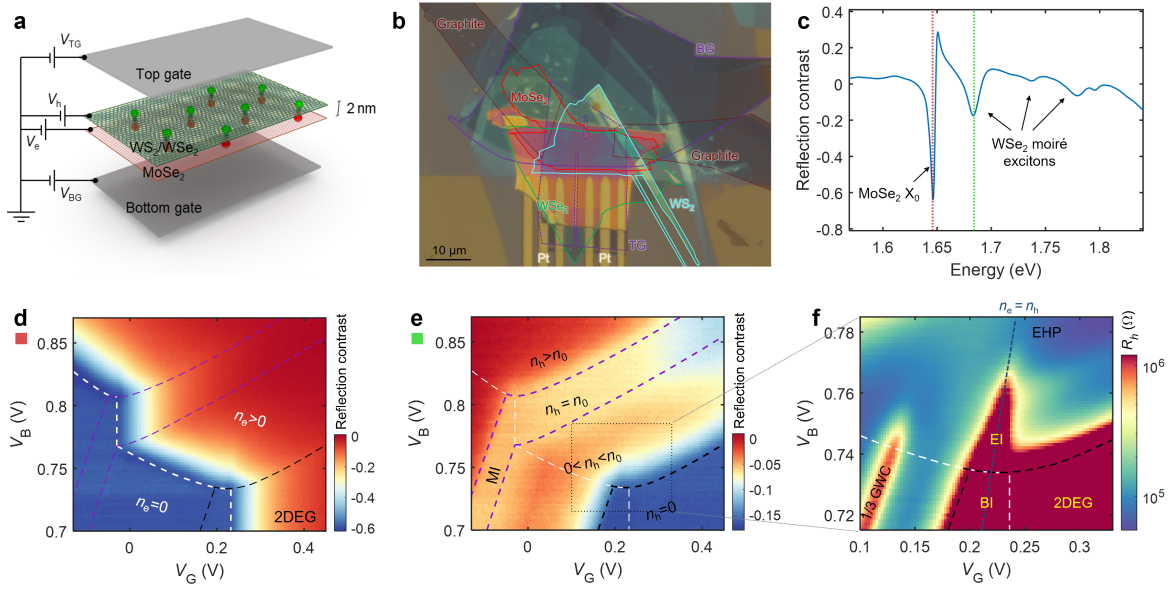

**Fig. 1 | Electron-hole bilayers with a moiré potential.**

**a**, Schematic illustration of the e-h bilayer device with a WS$_2$/WSe$_2$ moiré heterostructure as the hole layer and a monolayer MoSe$_2$ as the electron layer. The electron and hole layers are separated by a 2 nm thin hBN spacer. The moiré superlattice is defined by the WS$_2$/WSe$_2$ layers with a period of ~8 nm (moiré density $n_0 = 1.8 \times 10^{12}$ cm$^{-2}$). Green and red spheres represent holes and electrons respectively.

**b**, Optical microscopy image of the fabricated device D1, showing the layer boundaries. Two graphite electrodes are made to contact the MoSe$_2$ layer. Multiple Pt electrodes contact the WS$_2$/WSe$_2$ layer.

**c**, Reflection contrast spectrum when the device is undoped, measured at $T = 2$ K. The intralayer exciton peak of MoSe$_2$ (X$_0$) and multiple moiré excitons of WSe$_2$ can be clearly identified.

**d**, Reflection contrast at the MoSe$_2$ X$_0$ peak energy (1.646 eV, red dotted line in **c**), as a function of $V_G$ and $V_B$. The vertical electric field is fixed at 0.38 V/nm; MoSe$_2$ layer is chosen as the ground ($V_e = 0$). The MoSe$_2$ X$_0$ peak is known to lose its oscillator strength quickly when the MoSe$_2$ layer is doped with electrons, allowing us to determine the doping boundaries for the MoSe$_2$ layer (white dashed line).

**e**, Reflection contrast at the WSe$_2$ main moiré exciton peak (1.684 eV, green dotted line in **c**), with doping boundaries outlined by the black dashed lines. The WS$_2$/WSe$_2$ is undoped in the bottom-right corner ($n_h = 0$), where the reflection contrast is strongly negative (blue). A stripe of increased reflection contrast is characteristic of the MI at $n_h = n_0$ (purple dashed lines).

**f**, Hole resistance $R_h$ in the WS$_2$/WSe$_2$ layer as a function of $V_G$ and $V_B$. The MoSe$_2$ layer is open-circuited, and standard four-terminal resistance is measured for the WS$_2$/WSe$_2$ layer. Apart from trivial insulating phases (BI and 2DEG) where $n_h = 0$, high resistance is observed in the triangular region along the net charge neutrality line (blue dashed line), which originates from the dipolar EI phase.

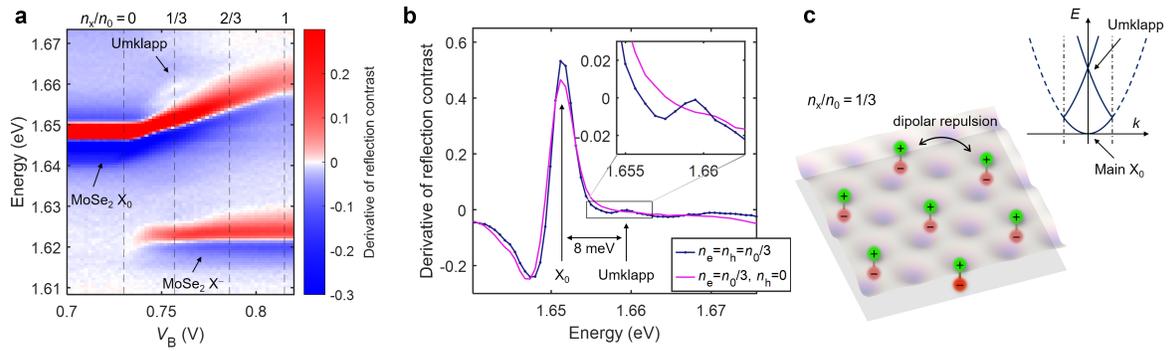

**Fig. 2 | Spectroscopic signatures of exciton crystallization.**

**a**, Reflection spectrum linecut along net charge neutrality (blue dashed line in Fig. 1f) as a function of the interlayer bias voltage $V_B$. A satellite peak corresponding to Umklapp scattering from the exciton crystal emerges at the exciton density $n_x (= n_e = n_h) = n_0/3$. The energy derivative of reflection contrast is plotted to visualize weak features.

**b**, Blue, reflection spectrum at $n_x = n_0/3$. The Umklapp scattering peak at higher energy is clearly present. Magenta, reflection spectrum at $n_e = n_0/3$ and $n_h = 0$. There is no Umklapp scattering peak with only electron doping.

**c**, Illustration of the exciton crystal at $n_x = n_0/3$ and the resulting Umklapp scattering mechanism. The exciton lattice folds the MoSe$_2$ X$_0$ dispersion (blue parabolic curve) back to the Brillouin zone center, giving rise to a higher-energy optical transition. Green and red spheres represent holes in the WS$_2$/WSe$_2$ layer and electrons in the MoSe$_2$ layer, respectively. Note that the exciton crystal is made of interlayer excitons, but the schematic exciton dispersion is for MoSe$_2$ intralayer excitons, which serves as a sensing probe.

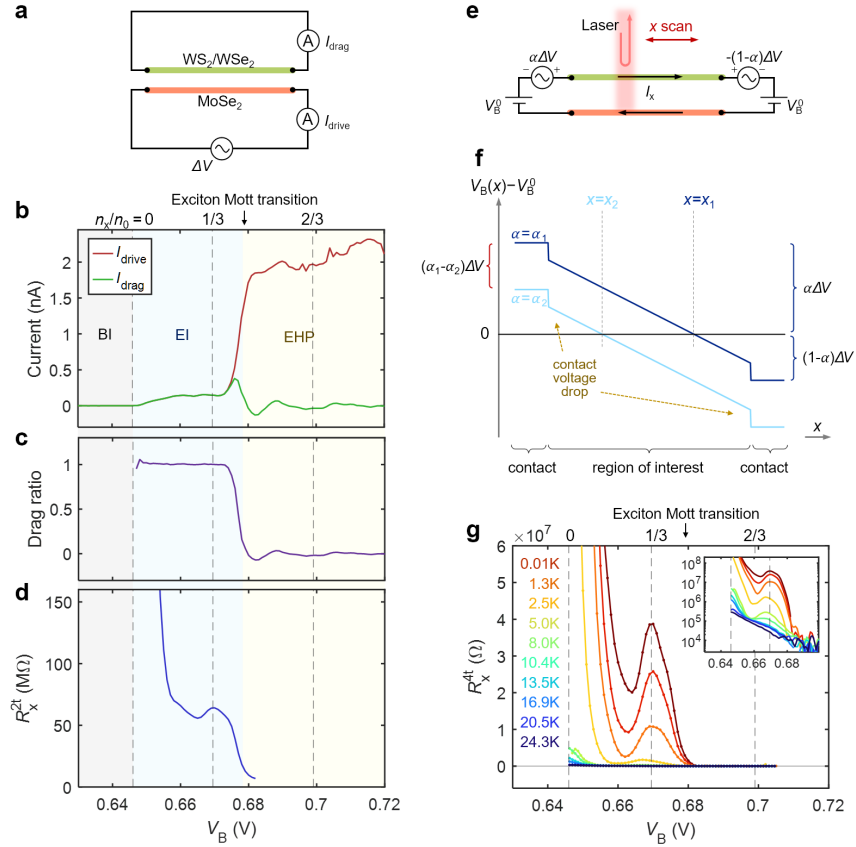

**Fig. 3 | Exciton transport behavior in a moiré potential.**

**a**, Schematic of the Coulomb drag measurement setup. An a.c. voltage excitation $\Delta V$ is applied to the electron layer (MoSe$_2$), generating a drive current $I_{\text{drive}}$. The hole layer (WS$_2$/WSe$_2$) is short-circuited, while the induced drag current $I_{\text{drag}}$ is measured.

**b**, Measured drive current $I_{\text{drive}}$ and drag current $I_{\text{drag}}$ as functions of $V_B$ along the charge neutrality line $n_e = n_h$. A larger electric field (0.43 V/nm instead of 0.38 V/nm) is applied for better contact performance, which reduces the $V_B$ onset of exciton doping by 0.09 V.

**c**, Drag ratio $I_{\text{drag}}/I_{\text{drive}}$ obtained from **b**. In the BI phase (light grey) at low $V_B$, no current is observed due to the absence of any carrier. In the EI phase (light blue) at intermediate $V_B$, equal drive and drag currents appear, indicating perfect exciton drag. At large $V_B$, the drag response decreases quickly to almost zero. This corresponds to an exciton Mott transition into an EHP phase (light yellow) above $n_x \approx 0.44 n_0$.

**d**, Two-terminal resistance $R_x^{2t}$ as a function of $V_B$, defined as $\Delta V/I_{\text{drive}}$. In the perfect drag regime, it characterizes the two-terminal resistance of interlayer excitons. It shows a resistance peak centered at 1/3 filling, suggesting the formation of an exciton crystal.

**e**, Schematic of the optical-electrical four-terminal measurement setup. On top of a common interlayer bias voltage $V_B^0$, the interlayer bias on the left side is increased by $\alpha \Delta V$ while the bias on the right side is decreased by $(1 - \alpha)\Delta V$. The potential difference $\Delta V$ generates an exciton current $I_x$. The exciton potential distribution is then optically read out. A laser probe is scanned along the device while monitoring the reflected light intensity, which is sensitive to local $V_B$.

**f**, Schematic spatial mapping of the local $V_B$ along the device. At a given $\alpha = \alpha_1$, the laser probe is scanned along $x$ to find the position $x = x_1$ where the local $V_B$ is not affected by $\Delta V$ (dark blue curve). As $\alpha$ is changed, this special position changes accordingly (light blue curve). Measuring the relation between such $x$ and $\alpha$ gives the exciton potential gradient $\Delta V \cdot d\alpha/dx$, which is proportional to the exciton resistance $R_x$.

**g**, Optically measured four-terminal exciton resistance as a function of $V_B$, confirming a resistance peak at 1/3 exciton filling. Inset, same data but plotted in log scale.

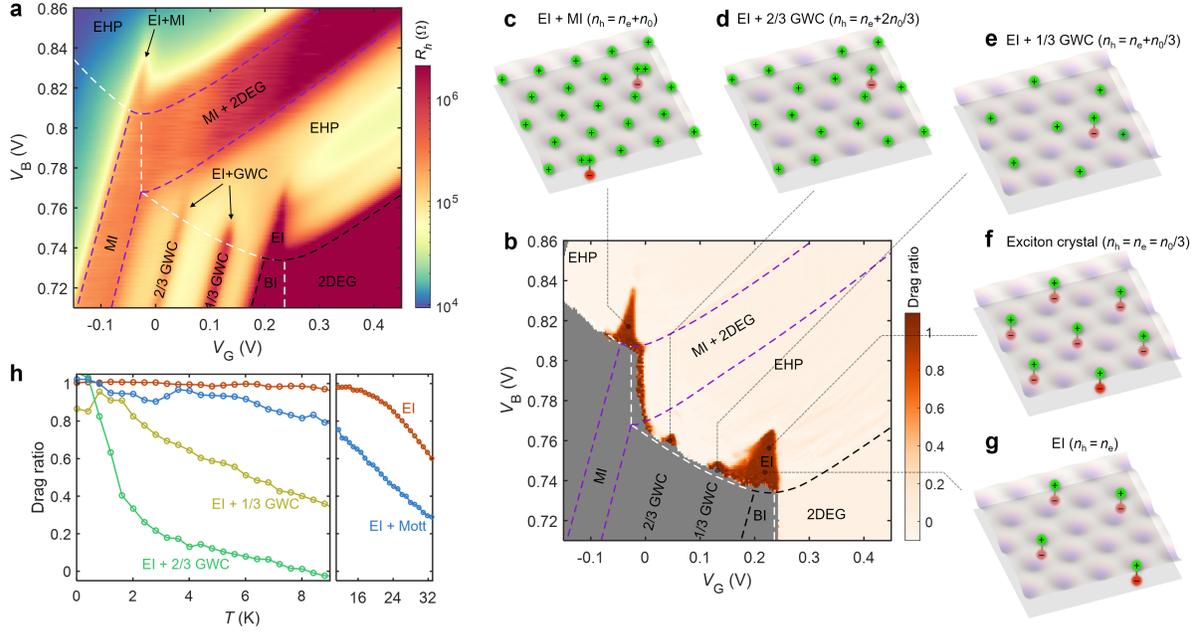

**Fig. 4 | Coexisting interlayer excitons and charge crystals.**

**a,** Hole-layer resistance $R_h$ as a function of $V_B$ and $V_G$ measured at $T = 2$ K and vertical electric field 0.38 V/nm. In addition to trivial insulating phases (BI and 2DEG) at $n_h = 0$, there are multiple correlated insulating phases despite finite $n_h$. Three of them appear in the $n_e = 0, n_h > 0$ region, including the MI state at $n_h/n_0 = 1$ and two GWC states at $n_h/n_0 = 1/3$ and $2/3$; four triangle-shaped insulating phases appear when holes and electrons are both doped and their density imbalance is 0 (the pure EI phase), $n_0$ (the EI+MI phase), $n_0/3$ or $2n_0/3$ (EI+GWC phases).

**b,** Closed-circuit drag ratio measured at lattice temperature $T = 0.01$ K. Besides the EI phase at $n_h = n_e$, EI+MI state and EI+GWC states on the left side of the charge neutrality also exhibit perfect Coulomb drag. The bottom-left region, where the MoSe$_2$ layer is not doped, is grayed out because $I_{drive} = 0$.

**c-g,** Schematic illustration of different correlated phases of the moiré e-h bilayer, including the EI+MI state (**c**), EI+GWC states (**d-e**), pure EI phase at equal electron-hole densities (**g**), and the exciton crystal state at 1/3 exciton filling within the EI phase (**f**).

**h,** Temperature-dependent drag ratios for different phases. The drag remains nearly perfect (ratio above 0.9) up to 21 K, 6 K, 2 K, and 0.7 K for the pure EI phase, the EI+MI phase, the EI+1/3 GWC phase, and the EI+2/3 GWC phase, respectively.

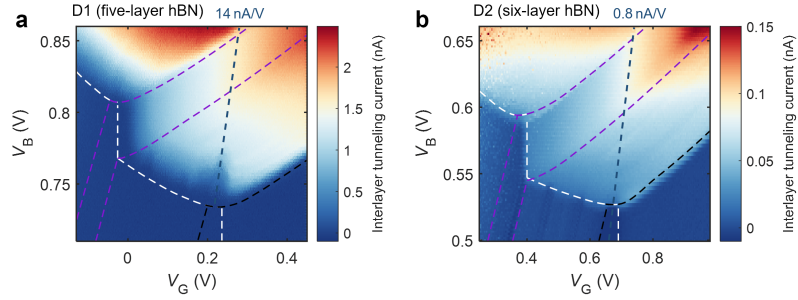

**Extended Data Fig. 1 | Interlayer tunneling current.**

**a**, Interlayer tunneling current for device D1 as a function of $V_G$ and $V_B$ at antisymmetric gating $V_{BG}/2 - V_{TG}/2 = 2.5$ V (electric field 0.38 V/nm). Fitting the tunneling current along the charge neutral line (blue dashed line) as a function of $V_B$ gives an increase rate of 14 nA/V, indicating a tunneling resistance of ~$10^8$ Ω.

**b**, Interlayer tunneling current for device D2 at antisymmetric gating $V_{BG}/2 - V_{TG}/2 = 3.5$ V (electric field 0.41 V/nm). A slightly thicker hBN barrier (6-layer hBN for D2, 5-layer hBN for D1) leads to an even larger tunneling resistance of ~$10^9$ Ω.

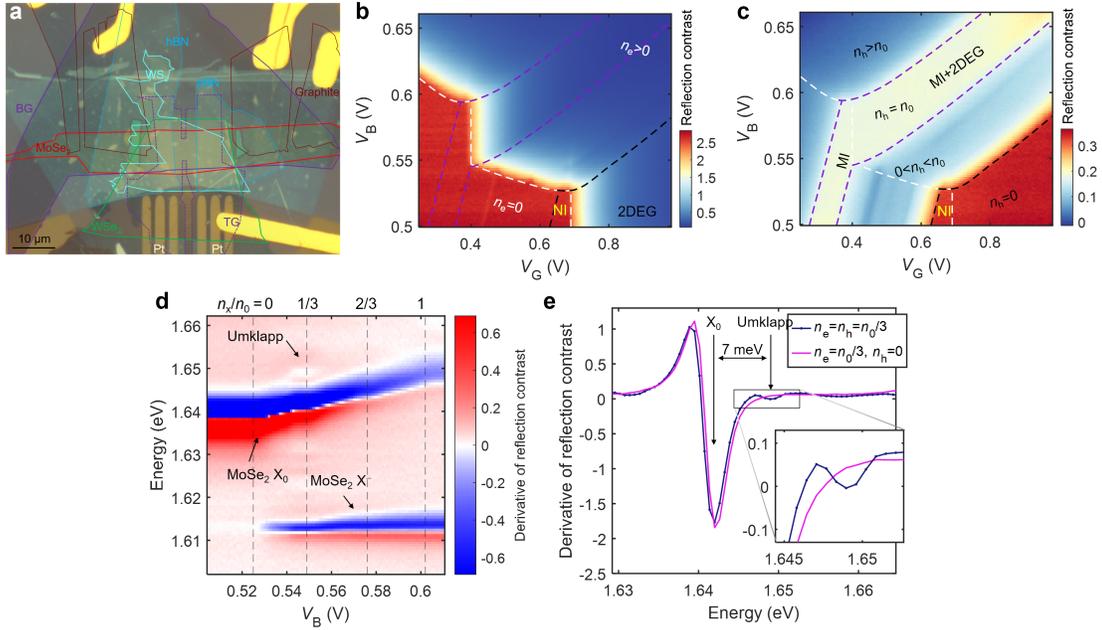

**Extended Data Fig. 2 | Main results from device D2.**

**a**, Optical microscopy image of device D2.

**b-c**, Reflection contrast intensity at the MoSe$_2$ main exciton peak (1.637 eV) and WSe$_2$ main moiré exciton peak (1.671 eV), respectively, as functions of $V_G$ and $V_B$. The vertical electric field is fixed at 0.41 V/nm. The general phase diagram is consistent with device D1 shown in Fig. 1.

**d**, Reflection spectrum linecut along net charge neutrality as a function of $V_B$, showing the emergence of a satellite peak (Umklapp scattering) at 1/3 exciton filling.

**e**, Blue, reflection spectrum at 1/3 exciton filling ($n_e = n_h = n_0/3$). Magenta, reflection spectrum when the system is at similar doping density but doped exclusively with electrons, not excitons ($n_e = n_0/3$ and $n_h = 0$).

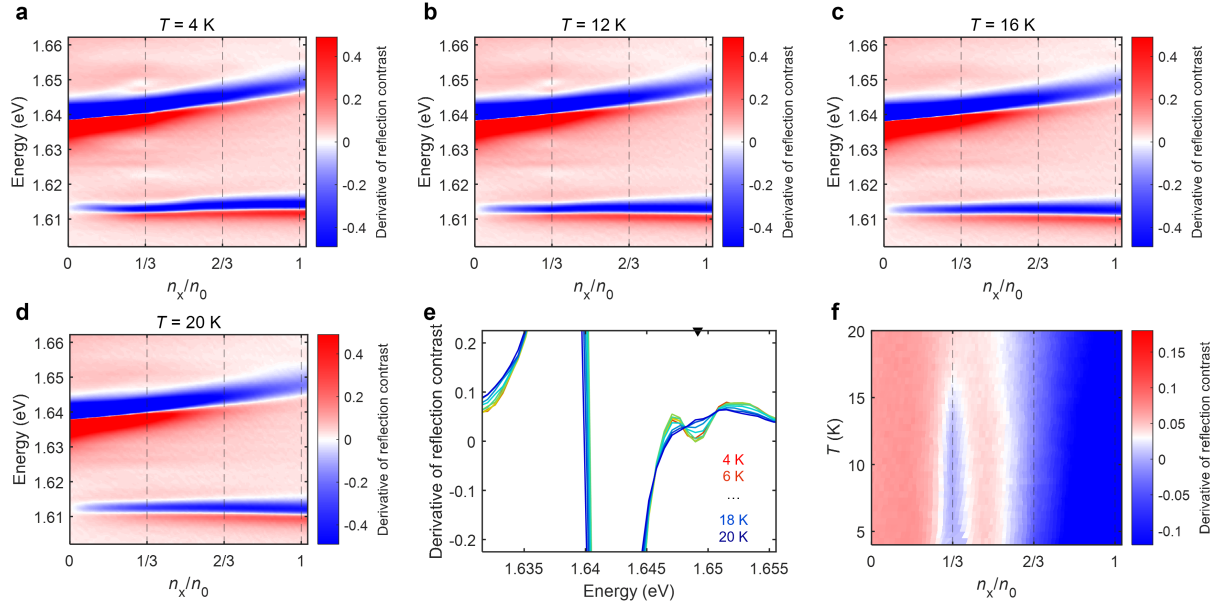

**Extended Data Fig. 3 | Temperature dependence of the Umklapp peak (device D2).**

**a-d**, Reflection spectra along the net charge neutrality line for different temperatures, as a function of filling factor $n_x/n_0$. The Umklapp peak at 1/3 filling is well defined at low temperatures, becomes weaker at 16 K, and eventually disappears at 20 K.

**e**, Evolution of the reflection spectrum at $n_x/n_0 = 1/3$ at various temperatures. The spectra at low temperatures ($T < 14$ K) all collapse together and clearly show the Umklapp peak, suggesting a well-defined exciton crystal. At higher temperatures it weakens, and becomes completely featureless at 20 K.

**f**, The spectrum intensity at 1.649 eV (black triangle in **e**), as a function of both exciton filling factor and temperature. The Umklapp peak appears only at 1/3 filling, and gradually diminishes after ~15 K.

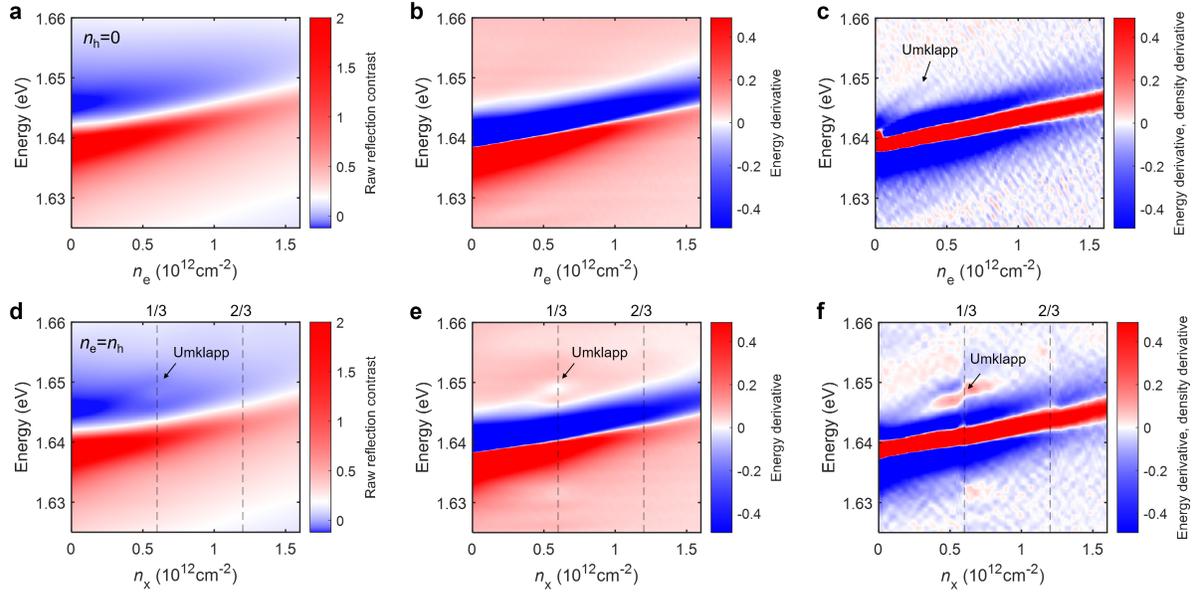

**Extended Data Fig. 4 | Umklapp peaks in electronic Wigner crystals and exciton crystals (device D2).**

**a**, Raw reflection contrast spectrum as a function of electron doping density when the hole layer is undoped. Only the main $X_0$ peak is visible.

**b**, Energy derivative of the spectra shown in **a**. Still only the main $X_0$ peak is visible.

**c**, Density derivative of the spectra shown in **b**. The Umklapp peak appears below density $\sim 3 \times 10^{11} \mathrm{cm}^{-2}$, consistent with spontaneous formation of electronic Wigner crystals when the density is sufficiently low. The double derivative suppresses the slow-decaying tail of the main peak, revealing the weak Umklapp peak.

**d**, Raw reflection contrast spectrum as a function of exciton doping density along net charge neutrality line. The Umklapp scattering peak at 1/3 exciton filling, although weak compared to the main peak, is readily visible without taking any derivative.

**e**, Energy derivative of the spectra shown in **d**. The Umklapp peak is very clear.

**f**, Density derivative of the spectra shown in **e**. The Umklapp peak in **d-f** is substantially stronger than the electronic Wigner crystal case, suggestive of a more ordered crystal structure. This can be explained by the presence of the moiré potential defining a common crystal orientation. In contrast, the electronic Wigner crystal in monolayer $MoSe_2$ does not have an intrinsically preferred orientation, which is usually determined by local defect or strain in $MoSe_2$ and therefore more vulnerable to sample inhomogeneities.

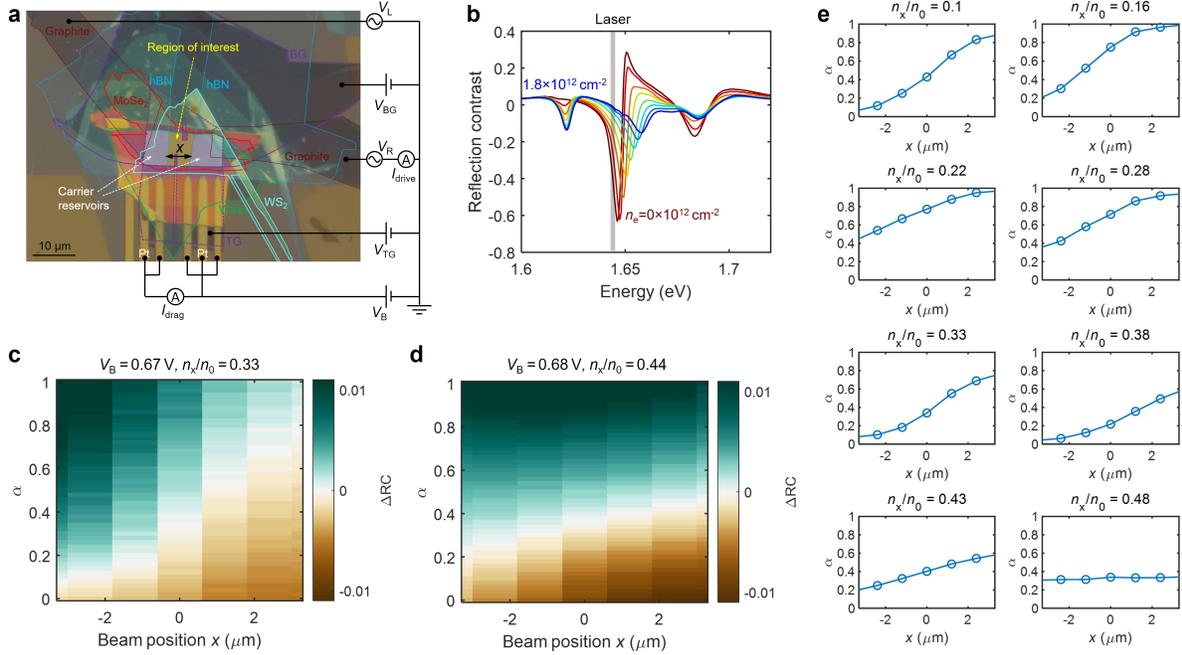

**Extended Data Fig. 5 | Hybrid electrical-optical measurement for exciton resistance.**

**a**, Circuit diagram of the measurement. The a.c. drive voltage is distributed on the two MoSe$_2$ contacts ($V_L = \alpha \Delta V$ and $V_R = -(1-\alpha)\Delta V$), with portion $\alpha$ on the left. It excites an exciton current in the EI, which is measured as $I_{\text{drive}}$. The potential distribution is then optically readout.

**b**, Density dependence of the reflection contrast spectrum. A monochromatic laser (gray line) reflected back from the device is used to probe the local potential distribution.

**c-d**, Measured a.c. optical response at two representative doping conditions. For exciton filling $n_x/n_0 = 1/3$ (**c**), the zero crossing point of the optical signal depends strongly on the position, indicating a large potential gradient due to the high resistance in the exciton crystal state. For exciton filling 0.44 (**d**), the resistance is much smaller.

**e**, $(\alpha, x)$ relation by tracing the zero crossing point of the a.c. optical signal.

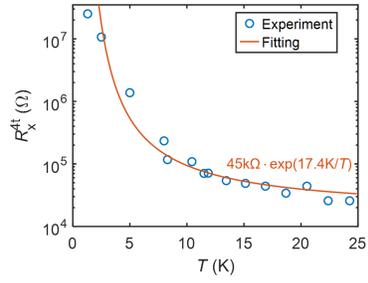

**Extended Data Fig. 6 | Thermal activation gap of the exciton crystal.**

Empty circles, optically measured $R_x^{4t}$ as a function of temperature at exciton filling $n_x/n_0 = 1/3$. Solid curve, fitting result using a thermal activation function $R_0 \exp(T_0/T)$.